\documentclass[aps,prb,preprint,superscriptaddress]{revtex4-1}

\usepackage{graphicx,amsmath,bm}

\usepackage{amssymb}
\usepackage{amsfonts}
\usepackage{mathtools}

\begin{document}


\title{Transition to collapsed tetragonal phase in CaFe$_2$As$_2$ single 
crystals as seen by $^{57}$Fe M\"ossbauer spectroscopy}

\author{Sergey L. Bud'ko}

\affiliation{Ames Laboratory, US DOE and Department of Physics and Astronomy, 
Iowa State University, Ames, Iowa 50011, USA}

\author{Xiaoming Ma}

\affiliation{Ames Laboratory, US DOE and Department of Physics and Astronomy, 
Iowa State University, Ames, Iowa 50011, USA}
\affiliation{Institute of Applied Magnetics, Key Laboratory for Magnetism and 
Magnetic Materials of the Ministry of Education, Lanzhou University, Lanzhou 
730000, Gansu Province, China}

\author{Milan Tomi{\'c}}

\affiliation{Institut f\"ur Theoretische Physik, Goethe-Universit\"at Frankfurt,
Max-von-Laue-Stra{\ss}e 1, 60438 Frankfurt am Main, Germany}

\author{Sheng Ran}

\altaffiliation{Currently at the Department of Physics, University of 
California - San Diego}
\affiliation{Ames Laboratory, US DOE and Department of Physics and Astronomy, 
Iowa State University, Ames, Iowa 50011, USA}

\author{Roser Valent\'\i}

\affiliation{Institut f\"ur Theoretische Physik, Goethe-Universit\"at Frankfurt,
Max-von-Laue-Stra{\ss}e 1, 60438 Frankfurt am Main, Germany}

\author{Paul C. Canfield}

\affiliation{Ames Laboratory, US DOE and Department of Physics and Astronomy, 
Iowa State University, Ames, Iowa 50011, USA}

\date{\today}

\begin{abstract}
Temperature dependent measurements of $^{57}$Fe M\"ossbauer spectra on  
CaFe$_2$As$_2$ single crystals  in the tetragonal and collapsed tetragonal 
phases are reported.  Clear features in the temperature dependencies of the 
isomer shift, relative spectra area and quadrupole splitting are observed at the 
transition from the tetragonal to the collapsed tetragonal phase. From the 
temperature dependent isomer shift and spectral area data, an average stiffening 
of  the phonon modes in the collapsed tetragonal phase is inferred. The 
quadrupole splitting increases by $\sim 25\%$ on cooling from room temperature 
to $\sim 100$ K in the tetragonal phase and is only weakly temperature dependent 
at low temperatures in the collapsed tetragonal phase, in agreement with the
anisotropic thermal expansion in this material. In order to gain microscopic 
insight about these measurements we perform {\it ab initio} density functional 
theory calculations of the electric field gradient and the electron density  of 
CaFe$_2$As$_2$ in both phases. By comparing the experimental data with the 
calculations we are able to fully characterize the crystal structure of the 
samples in the collapsed-tetragonal phase through determination of the  As 
$z$-coordinate. Based on the obtained 
temperature dependent structural data we are able 
to propose charge saturation of the Fe - As bond region as the mechanism behind the 
stabilization of the collapsed-tetragonal phase at ambient pressure. 
\end{abstract}


\maketitle

\section{Introduction}

CaFe$_2$As$_2$ is conceivably one of the most interesting materials in the 
family of Fe-As based superconductors with a very fragile ground state and 
extreme sensitivity to the external pressure and/or strain. At ambient pressure, 
a clear, sharp, first order transition at $\sim 170$ K from high temperature 
tetragonal -  paramagnetic phase to low temperature orthorhombic -  
antiferromagnetic phase  was  observed in thermodynamic, transport and 
microscopic measurements, in single crystals grown out of Sn flux. \cite{nin08a, 
gol08a, wug08a, ron08a, can09a, alz11a} Under pressure, this transition is 
initially suppressed at a rate of 10-15 K/kbar \cite{can09a, tor08a, kre08a, 
gol09a, yuw09a, lee09a}. As the pressure increases, at about 3 kbar, a 
paramagnetic - collapsed tetragonal (cT) phase terminates the lower pressure 
orthorhombic - antiferromagnetic phase near 100 K. The cT transition temperature 
increases under pressure and reaches 300 K at $\sim 15$ kbar. \cite{tor08a, 
kre08a, gol09a} The uniaxial pressure derivatives of the structural/magnetic 
transition are large and of different signs. \cite{bud10a} The pressure - 
temperature phase diagram for CaFe$_2$As$_2$ is very sensitive to 
non-hydrostaticity of the applied pressure. \cite{can09a, yuw09a, tor09a, 
pro10a}

This extreme sensitivity of pure, as well as substituted CaFe$_2$As$_2$ single 
crystals to external stresses and strains yields an unexpected result: by a 
judicious choice of the annealing temperature and time (e.g. by control of  
internal stress fields via nanoscale precipitates) we can tune the 
structural/magnetic transition temperature in FeAs grown  CaFe$_2$As$_2$ 
crystals  in a manner similar to application of pressure: suppressing the 
antiferromagnetic/orthorhombic transition by over 70 K and furthermore, we can 
obtain the crystals with the cT ground state at ambient pressure.~\cite{ran11a} 
By adding transition metal substitution as an extra parameter, we can tune the 
ground state of the Ca(Fe$_{1-x}$T$_x$)$_2$As$_2$ (T = Co, Ni, Rh) crystals to 
be orthorhombic - antiferromagnetic, tetragonal - superconducting, tetragonal - 
paramagnetic, and collapsed tetragonal - paramagnetic, all at ambient pressure. 
\cite{ran12a,ran14a} Pressure measurements on Co-substituted CaFe$_2$As$_2$ 
further illustrate the similar effects of applied pressure and post - growth 
annealing and quenching of samples \cite{gat12a} and also underscore the 
system's pressure sensitivity with $d T_{(AFM/ortho)}/dP \approx -110$  K/kbar 
and $d T_c/dP \approx -6$ K/kbar.

The aforementioned tunability of  CaFe$_2$As$_2$ created the opportunity for 
studies of the cT phase at ambient pressure (e.g. in FeAs grown CaFe$_2$As$_2$ 
single crystals quenched from 960$^{\circ}$ C), using experimental techniques that are either 
difficult or impossible to combine with hydrostatic pressure. 
\cite{soh13a,fur14a,bud13a,dha14a,gof14a,sap14a}

$^{57}$Fe M\"ossbauer spectroscopy has been widely used to study Fe-As based 
superconductors and related materials, \cite{rot08a,now09a,now10a,nat13a,jas15a} 
including Sn-grown CaFe$_2$As$_2$ single crystals \cite{alz11a,kum09a, liz11a} 
and powdered single crystals \cite{bla11a}  primarily for probing magnetism (although it 
should be noted that grinding of CaFe$_2$As$_2$ affects the 
macroscopic and microscopic magnetic properties \cite{max15a}). In this work we 
will use this local probe technique to perform a detailed temperature-dependent 
study of the cT transition (and the cT phase) in single crystals of undoped,  FeAs-grown
960$^{\circ}$ C quenched \cite{ran11a},  CaFe$_2$As$_2$. In these  CaFe$_2$As$_2$ samples  
both the high temperature tetragonal and low temperature cT phases are paramagnetic, with no magnetic order. We will compare the 
results with the available literature data on temperature and pressure induced 
transition to the cT phase. Some initial $^{57}$Fe M\"ossbauer spectroscopy data 
on very similar FeAs-grown CaFe$_2$As$_2$ crystals with cT ground state were 
published in Ref. \onlinecite{ran11a} and the pressure induced transition to the 
cT phase at room temperature in Sn- grown sample was recently studied by  
M\"ossbauer spectroscopy in Ref. \onlinecite{sak14a}. 

The isomer shift and the quadrupole splitting observed in M\"ossbauer 
spectroscopy can be directly related to the electron density at the nucleus and 
the electric field gradient respectively, both of which can be calculated in the 
framework of \emph{ab-initio} Density Functional Theory. We will employ these 
calculations in order to provide a theoretical interpretation of the 
measurements, as well as to obtain a full structural characterization of the 
960$^{\circ}$ C quenched samples in the collapsed-tetragonal phase by 
determining the As $z$-coordinates. We will also analyze the temperature dependent 
structural data in order to understand the microscopic mechanism responsible for 
the stabilization of the collapsed-tetragonal phase at ambient pressure. 

\section{Experimental Details and Computational Methods}

\begin{figure}
\begin{center}
\includegraphics{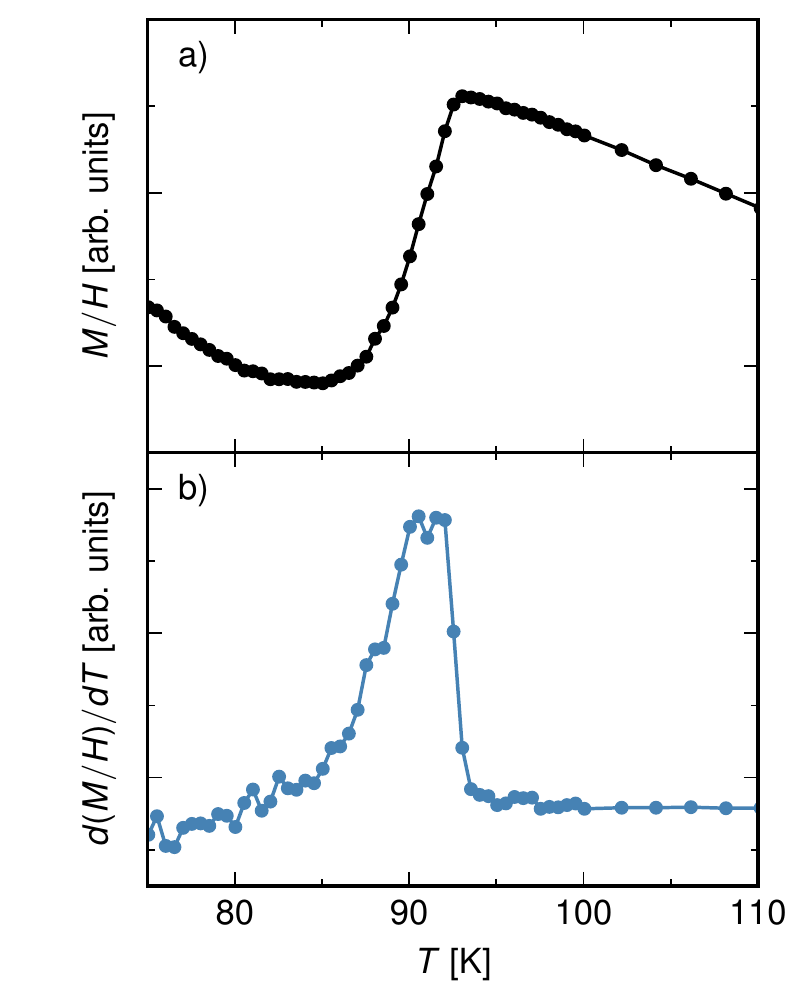}
\end{center}
\caption{Temperature dependent magnetization, $M/H(T)$ and its temperature 
derivative, $d(M/H)/dT$, of a part of CaFe$_2$As$_2$ crystals mosaic used for 
M\"ossbauer measurements in the vicinity of the cT transition. Note that the 
paper and adhesive of the mount give some small, weakly temperature dependent contribution to $M/H(T)$. 
}\label{MT}
\end{figure}

CaFe$_2$As$_2$ single crystals were grown from ternary melts rich in FeAs, 
following the procedure described in Ref.  \onlinecite{ran11a}. The excess 
liquid was decanted at 960$^{\circ}$ C, essentially quenching the samples from  
960$^{\circ}$ C to room temperature. The cT transition temperature of the sample 
was $\sim 90$ K as confirmed by dc magnetization measurements (see Fig. 
\ref{MT}). \cite{ran11a} In the rest of the paper we will  use CaFe$_2$As$_2$  
to refer to these FeAs grown, CaFe$_2$As$_2$  samples quenched from 960$^{\circ}$ C to 
room temperature.

M\"ossbauer spectroscopy measurements were performed using a SEE Co. 
conventional, constant acceleration type spectrometer in transmission geometry 
with  a $^{57}$Co(Rh) source kept at room temperature. The absorber was 
prepared as a mosaic of   0.04 - 0.1 mm thick  single crystals held between two 
paper disks by a small amount of a  pressure sensitive adhesive. 
The gaps between the individual crystals were kept as small as 
possible. The mosaic had the $c$ - axis perpendicular to the disks (with the 
accuracy of or better than $\pm 10^{\circ}$) and arbitrary in-plane orientation.  
Since CaFe$_2$As$_2$ and related materials are known to be very sensitive to the 
stresses and strains, including those caused by sample mounting, \cite{ran14b} 
we performed dc magnetization measurements on a subset of the mosaic with the 
same mounting as for the M\"ossbauer measurements (Fig. \ref{MT}). The cT 
transition is still sharp, without apparent shift of the transition temperature.

The constrains on the thickness of the CaFe$_2$As$_2$  crystals that can be cleaved without signficant mechanical damage result in the absorber having 5-20 mg natural Fe per cm$^2$ that corresponds to a "thick" absorber limit. \cite{gut11a} The consequences of using "thick"  absorber include thickness broadening of the absorption lines, increased error bars in absolute determination of the quadrupole splitting and different degrees of saturation of the absorption lines of different intensity. The intrinsic linewidth found in inorganic compounds is often in the range of 0.26 - 0.28 mm/s, \cite{gut11a} so the broadening observed in our measurements is not extremely severe. The other effects of the absorber thickness manifest as "smooth" contribution to the temperature dependent data and cannot cause the sharp features observed in the temperature dependences of the hyperfine parameters at the cT transition.

The absorber holder consists of two nested white Delrin cups.  The absorber 
holder was locked into thermal contact with a copper block with a temperature 
sensor and a heater, and was aligned with the $\gamma$ - source and detector.  
The  $c$-axis of the crystals in the mosaic was parallel to the M\"ossbauer 
$\gamma$ beam. The absorber was cooled to a desired temperature using a Janis 
model SHI-850-5 closed cycle refrigerator (with vibration damping). 
The measurements were performed on warming, after initially 
cooling the absorber down to a base temperature of 4.6 K. The driver 
velocity was calibrated using an $\alpha$-Fe foil and all isomer shifts (IS) are 
quoted relative to the $\alpha$-Fe foil at room temperature. All the M\"ossbauer 
spectra were fitted by the commercial software package MossWinn. \cite{MW}

{\it Ab initio} Density Functional Theory (DFT) calculations were performed with 
the scalar relativistic Full Potential Local Orbital method 
(FPLO)\cite{Koepernik1999}. The electric field gradients were obtained according 
to the technique described in Ref. \onlinecite{Koch2010} and  implemented in the 
FPLO code. The Perdew-Burke-Erzenhof \cite{Perdew1996} exchange-correlation 
potential was employed for all calculations, and we considered  $20 \times 20 
\times 20$ k-point meshes, with charge density converged within $10^{-8}$ 
a.u.$^{-3}$. In order to calculate  the electron density at the location of atomic 
nuclei, we  assumed the model of a finite-size nucleus with  constant charge 
density. Consistency of the calculated electric field gradients and electron 
densities was verified against values obtained by the Full-Potential Linearized 
Augmented Plane-Wave method implemented in the WIEN2K code \cite{Blaha2001}.

\section{Experimental Results and Discussion}

\begin{figure}
\begin{center}
\includegraphics{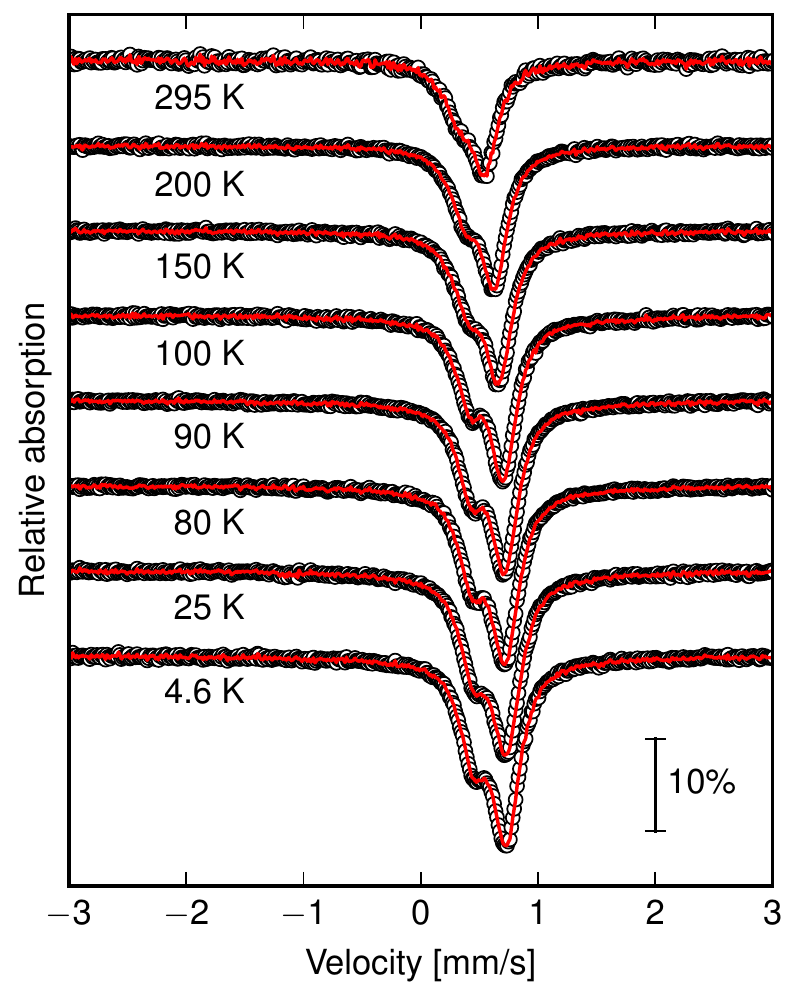}
\end{center}
\caption{(Color online) $^{57}$Fe M\"ossbauer spectra of CaFe$_2$As$_2$ at 
selected temperatures. Symbols - data, lines - fits. $T_{cT} \sim 90$ K, as shown in Fig. \ref{MT}.}\label{F1}
\end{figure}

The $^{57}$Fe M\"ossbauer spectra of CaFe$_2$As$_2$ at selected temperatures 
between the room temperature and the base temperature of 4.6 K are shown in 
Fig. \ref{F1}. At each temperature the spectrum is a quadrupole split doublet 
with the higher velocity line having almost twice the intensity of the lower 
velocity line. No secondary phase signal was observed in the spectra.  Visually, 
the temperature induced changes, even near $T_{cT} \sim 90$ K, are subtle.

\begin{figure}
\begin{center}
\includegraphics{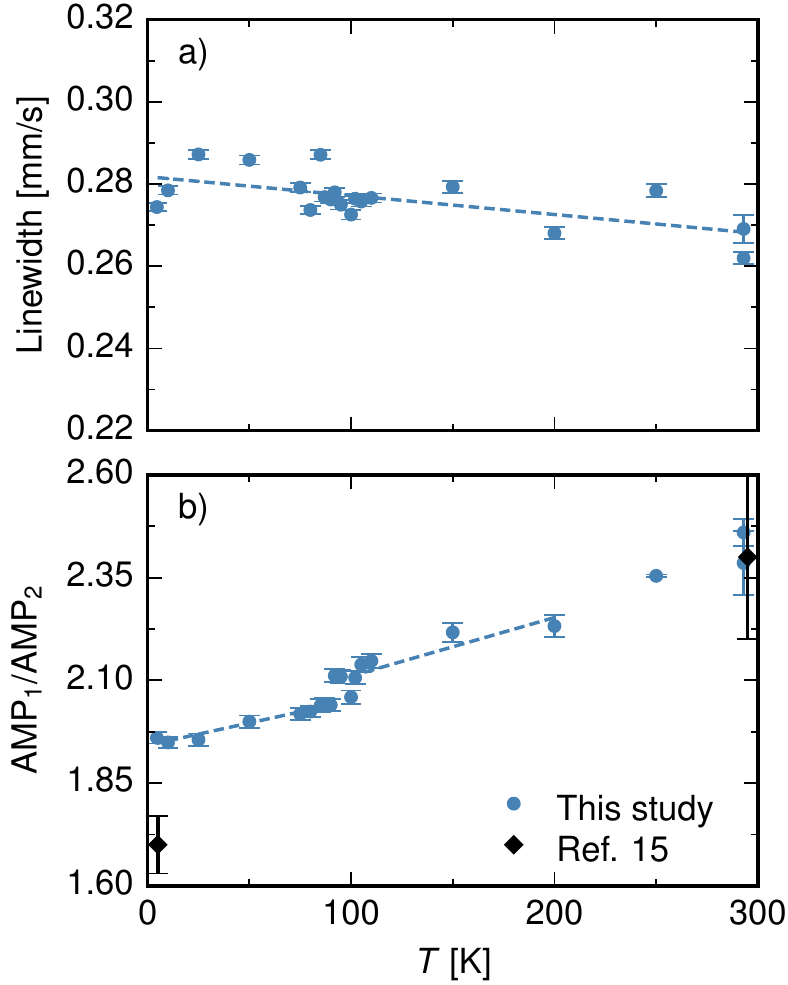}
\end{center}
\caption{(Color online) CaFe$_2$As$_2$ $^{57}$Fe M\"ossbauer data:  (a) Temperature dependence of the linewidth; (b) The 
intensity ratio of the doublet lines as a function of temperature. Circles - this work, rhombuses - Ref. \onlinecite{ran11a}. Dashed lines are linear fit to the data [panel a)] and guides to the eye [panel b)]. }\label{F2}
\end{figure}

The temperature dependent linewidth of the spectra is shown in Fig. \ref{F2} a). 
Overall  the linewidth increases by a few percent on cooling from 
room temperature to the base temperature. As has been discussed in Ref. 
\onlinecite{ran11a}, the point symmetry and the location of the Fe site in the 
ThCr$_2$Si$_2$ structure constrains the principal axis of the local electric 
field gradient tensor to the $c$ - crystalline axis, as a result, a doublet 
lines intensity ratio of  3 : 1 is expected for the mosaic with the $c$ - axis 
parallel to the $\gamma$ beam. Instead, the observed room temperature ratio is 
$\sim 2.4 : 1$ and it decreases to $\sim 1.95 : 1$ at the base temperature (Fig. 
\ref{F2} (b)).  The general 
behavior is consistent with the two temperature points reported earlier. 
\cite{ran11a} Several possible reasons for the doublet lines  intensity ratio 
being different from 3 : 1 were discussed in Ref. \onlinecite{ran11a}.  
Most probably, an intensity ratio
smaller than 3 : 1  and its 
further decrease on cooling may be due to the "thick" absorber
 conditions of the 
measurements (consistent with the large absorption observed (Fig. \ref{F1}).
We observe a small but distinct feature in the intensity ratio at the
transition temperature around $\sim 90$ K. The 
change in the intensity ratio at the cT transition reflects an increase of the 
degree of saturation (caused by the thick absorber) 
of the stronger spectral line, 
thus pointing to stiffening of the lattice in the low temperature cT phase.

\begin{figure}
\begin{center}
\includegraphics{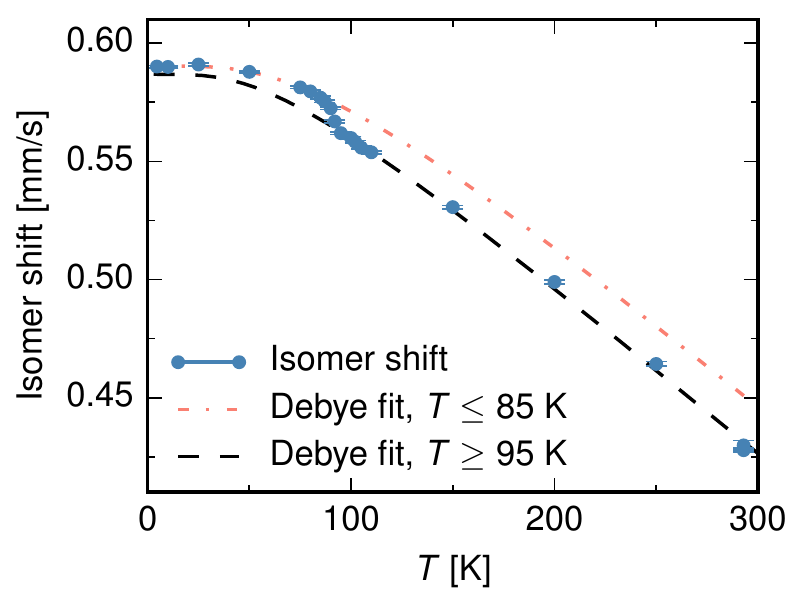}
\end{center}
\caption{(Color online) Temperature dependent isomer shift of  CaFe$_2$As$_2$. 
Dashed line - Debye fit for $T \ge 95$ K resulting in $\Theta_D = 239 \pm 16$ 
K, dash-dotted line Debye fit for $T \le 85$ K resulting in $\Theta_D = 316 \pm 
7$ K}\label{IS}
\end{figure}

In Fig.~\ref{IS} we show the measured  isomer shift (IS) which increases upon 
cooling  and has a clear step-like change between 95 K and 85 K. The isomer 
shift includes contributions from both the chemical shift and the second-order 
Doppler shift. The latter is known to increase convexly upon decreasing 
temperature, due to gradual depopulation of the excited phonon states. However, 
it should be constant at low temperature, because of the quantum mechanical 
zero-point motion. The chemical shift should not depend on temperature. The main 
contribution to the temperature dependence of the isomer shift is therefore from 
the second-order Doppler shift, and is usually described by the Debye model: 
\cite{gut11a,Greenwood1971} \\
\begin{equation}\label{eq:IS_Debye}
\delta(T)=\delta(0)-\frac{9}{2}\frac{k_BT}{Mc}\left(\frac{T}{\Theta_D}
\right)^3\int_0^ { \Theta_D/T }
\frac{x^3dx}{e^x-1},
\end{equation}
\\
where $c$ is the velocity of light,  $M$ is the mass of the $^{57}$Fe nucleus, 
and $\delta(0)$ is  the temperature-independent part. For the isomer shift data 
in Fig.~\ref{IS} separate Debye fits, for the tetragonal phase ($T \geq 95$ K) 
and for the collapsed tetragonal phase ($T \leq 85$ K) were performed. The fits 
yield $\Theta_D = 239 \pm 16$ K for the high temperature tetragonal phase and 
$\Theta_D = 316 \pm 7$ K for the low temperature cT phase. These results suggest 
that the lattice is stiffer in the cT phase. It has to be noted 
that the results of the fit in the low temperature cT phase should be taken with 
caution, since the fit is performed on a limited dataset with only minor 
temperature dependence. Still the conclusions are consistent with the 
rest of the data 
discussed in this work.

\begin{figure}
\begin{center}
\includegraphics{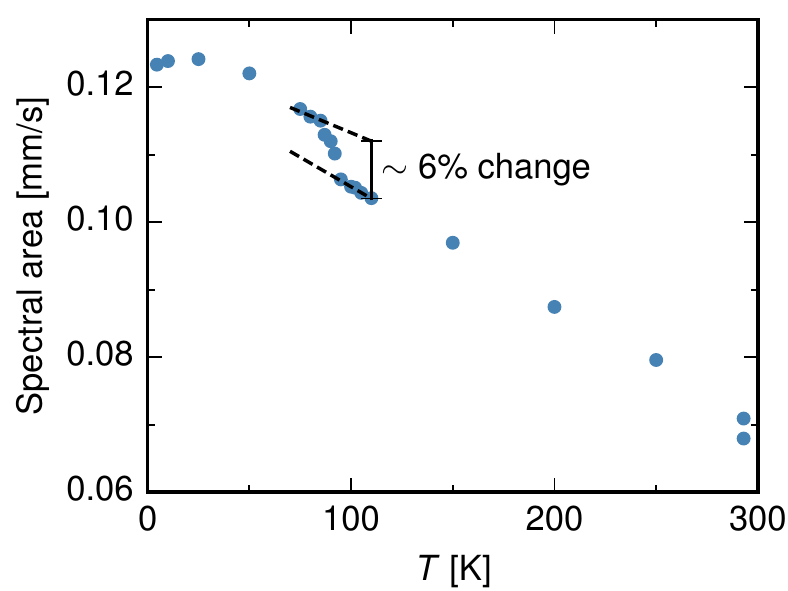}
\end{center}
\caption{(Color online) Temperature dependent relative spectral area of  
CaFe$_2$As$_2$. Dashed lines are guides to the eye. 6\% change is shown as a vertical bar.}\label{SA}
\end{figure}

The relative spectral area (Fig. \ref{SA}) also shows a distinct step - like 
feature in the region of the structural transition. Although we 
do not perform a Debye fit of the temperature dependent spectral area due to 
the possible effects of the thick absorber, the step - like increase of spectral area (by $\sim 6\%$) at the 
cT transition points to an increase of the Debye temperature, i.e. stiffening of 
the lattice, in the cT phase. Somewhat higher, $\sim 9\%$, increase in the relative spectral area is expected using $\Theta_D$ values from fits of temperature-dependent isomer shift and the Debye model for the recoil-free fraction: \cite{gut11a,Greenwood1971}
 \\
\begin{equation}
 f = \exp\left\{ \frac{-3E_{\gamma}^2}{k_B\Theta_DMc^2} \left[
  \frac{1}{4} + \left(\frac{T}{\Theta_D}\right)^2 
  \int_0^{\Theta_D/T}\frac{xdx}{e^x-1} \right]
 \right\},
\end{equation}
\\
(where $f$ is the recoil-free fraction, and E$_{\gamma}$ is the $\gamma$-ray energy) leading to the conclusion that the value of $\Theta_D$ in the cT phase may be slightly overestimated.

 Stiffening of the lattice upon transition to a collapsed tetragonal 
phase (under pressure) was clearly observed in BaFe$_2$As$_2$~\cite{mit11a} via 
compressibility measurements, however similar sets~\cite{gol09a,mit11a} of structural 
data for CaFe$_2$As$_2$ appear to be less convincing. Nonetheless,
 a detailed study 
of phonon spectra in CaFe$_2$As$_2$~\cite{mit10a} concluded that the phonons 
polarized in the $ab$ plane are very similar in the tetragonal and cT phases, 
whereas transverse acoustic phonons propagating along the $c$ direction stiffen 
very significantly in the cT phase.  These results are consistent with  
our analysis of the changes in the M\"ossbauer spectra
at the structural transition.

\begin{figure}
\begin{center}
\includegraphics{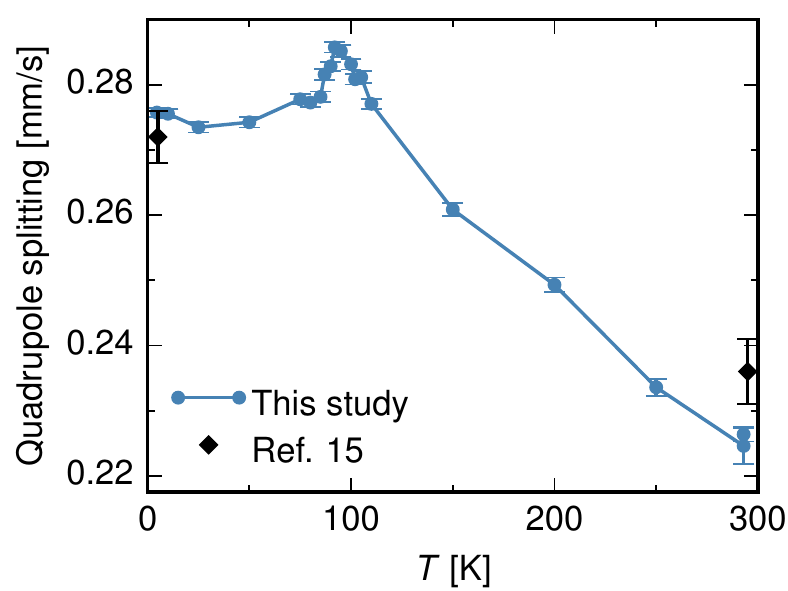}
\end{center}
\caption{(Color online) Temperature dependent quadrupole splitting of  
CaFe$_2$As$_2$. Circles - this work, rhombuses - Ref. 
\onlinecite{ran11a}.}\label{QS}
\end{figure}

The temperature dependent quadrupole splitting (QS)  data is presented in Fig. 
\ref{QS}. The total change of the QS with temperature is  $\sim 25 \%$. QS 
varies strongly in the tetragonal phase and is almost constant in the cT phase. 
The cT transition is clearly seen as a cusp in QS(T). Qualitatively, this 
behavior is consistent with the anisotropic thermal expansion of CaFe$_2$As$_2$ 
\cite{bud13a}:  the quadrupole splitting is related to the $c/a$ ratio, that 
changes considerably above the cT transition and significantly less in the cT 
phase. In a very general sense this behavior is consistent with the results from 
the Ref. \onlinecite{sak14a}  (with pressure as a control parameter): QS 
increases on approaching the cT transition, displays a cusp at the transition, 
and does not change in a significant manner in the cT phase. 
The origin of the cusp in the QS(T) data at the cT transition 
(possibly observed as a function of pressure \cite{sak14a} as well) is
however not clear 
and requires additional studies beyond the scope of the present work.

\section{Miscroscopic Modelling}

To better understand the physics behind the experimental observations we rely 
on the fact that the isomer shift and the quadrupole splitting are directly 
related to the electronic charge densities and the electric field gradients, 
which can be calculated \emph{ab-initio} in the framework of density
functional theory (DFT).

The isomer shift $\delta$ can be expressed~\cite{Greenwood1971} as:
\\
\begin{equation}\label{eq:isomer_shift}
 \delta = \frac{Ze^2}{5\varepsilon_0}(R_e^2-R_g^2)[\rho_a(0)-\rho_s(0)]
\end{equation}
\\
where $Z$ is the nuclear charge, $e$ is the elementary charge, $\varepsilon_0$ 
is the permittivity in vacuum, $R_g$ and $R_e$ are the nuclear ground and 
excited state radii and $\rho_s(0)$ and $\rho_a(0)$ are electron densities at 
the source and absorption nuclei respectively. The only quantity that varies in 
the measurement is the  electron density around the absorption nucleus 
$\rho_a(0)$ so that we can simply write Eq.~\eqref{eq:isomer_shift} as a linear 
relation
\\
\begin{equation}\label{eq:isomer_shift_simple}
 \delta = a[\rho_a(0)-b]
\end{equation}
\\
Note that for $^{57}$Fe the constant $a$ is negative \cite{eri89a, Lovell2002, 
Neese2002, Sinnecker2005, Han2006}. With this taken into account, part of the observed 
step-like increase in the isomer shift at the onset of the volume collapse (Fig. 
\ref{IS}) would correspond to the likewise drop in the electron density at the 
iron nucleus. This can be explained as an immediate consequence  of Fe charge 
delocalization brought in by the volume collapse as a result of a strong, first 
order, structural phase transition.

Furthermore, the quadrupole splitting $\Delta$ can be expressed as 
\cite{Greenwood1971}:
\\
\begin{equation}\label{eq:quadrupole_splitting}
 \Delta = \frac{1}{2}eQV_{zz}\sqrt{1+\frac{\eta^2}{3}}
\end{equation}
\\
where $Q$ is the nuclear quadrupole moment, $\eta=|V_{xx}-V_{yy}|/|V_{zz}|$ is 
the electric field gradient asymmetry, and  $|V_{xx}| \leq |V_{yy}| \leq 
|V_{zz}|$ are the eigenvalues of the electric field gradient tensor. Since the 
sample is tetragonal, the asymmetry $\eta$ is zero and we obtain the following 
linear equation
\\
\begin{equation}\label{eq:quadrupole_splitting_simple}
 \Delta = cV_{zz}
\end{equation}
\\
where $c=eQ/2$.

Understanding the origin of the QS behavior (Fig. \ref{QS}) based on 
Eq.~\ref{eq:quadrupole_splitting_simple} is quite involved since the electric 
field gradient $V_{zz}$ strongly depends on crystal structural details, in 
particular the details of the Fe-As bonding. 

The crystal structure of the tetragonal CaFe$_2$As$_2$ is that of ThCr$_2$Si$_2$ 
\cite{Hoffmann1985} and is described by the space group $I4/mmm$. The Wyckoff 
positions corresponding to Ca and Fe atoms are fully fixed by the symmetry, 
while the Wyckoff position corresponding to the As atoms has only $x$ and $y$ 
coordinates fixed, leaving the $z$-coordinate as the sole internal structural 
degree of freedom.

The structure of the  960$^{\circ}$ C quenched CaFe$_2$As$_2$ sample was 
characterized in the 5-300 K temperature range by X-ray diffraction measurements 
\cite{ran11a}. However, these measurements were able to determine only the unit 
cell parameters, and the As $z$-coordinate was left unknown, rendering the 
crystal structure information incomplete.

\begin{figure}
\begin{center}
\includegraphics{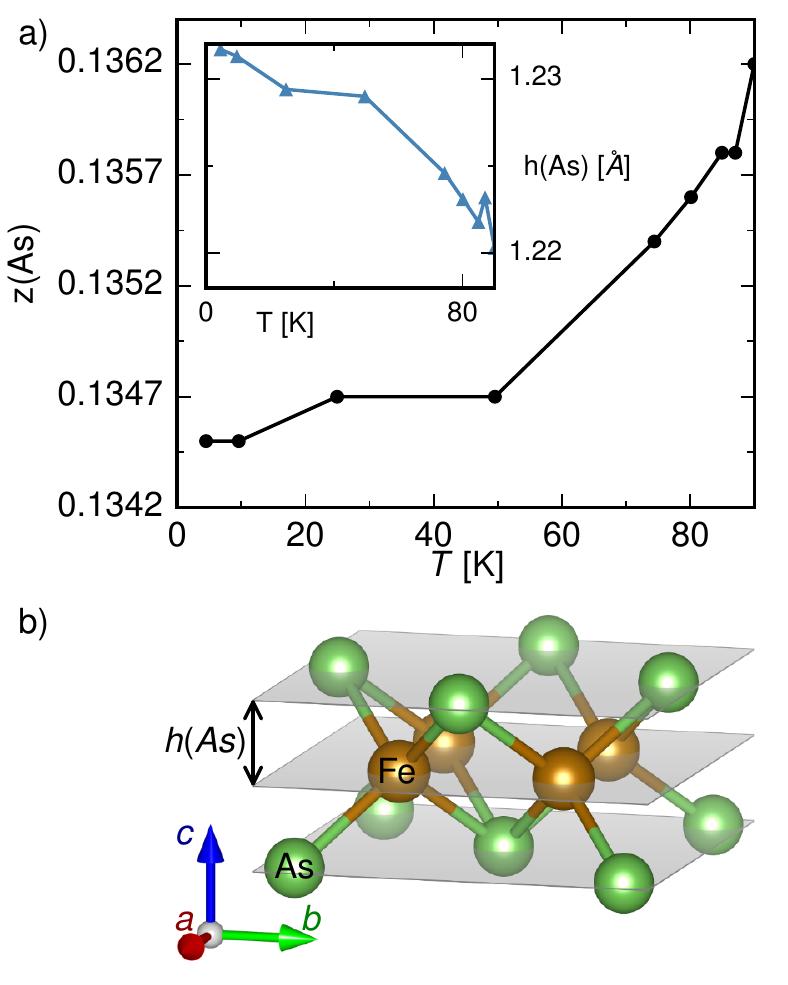}
\end{center}
\caption{(Color online) (a) Calculated temperature dependence of As 
$z$-coordinate in the collapsed-tetragonal phase. The inset shows temperature 
dependence of height of As atoms above the Fe plane. (b) Diagram showing 
geometry of the FeAs layer.}\label{fig:as_zh}
\end{figure}

Full crystal structure information is the starting point for any DFT 
calculation. Since we 
don't have here information of the As $z$-coordinate but we do have valuable 
information from experiment on isomer shifts and quadrupole splittings, we are 
going to perform a ``reverse engineering'', namely from the experimental data 
we will deduce the As $z$ coordinates as a function of temperature. We would 
like to point out that the task of providing an accurate determination of the 
As $z$-coordinates  has additional importance through the fact that As atom 
positions are controlling parameters in the tetrahedral coordination geometry of 
the Fe atoms (see Fig. \ref{fig:as_zh} b)), and they are thus a 
very sensitive driver of physics of the iron-arsenide layer \cite{Stewart2011}.

Obtaining the As $z$-coordinate in a straightforward way, by structural 
optimization within DFT, may be rather inaccurate because it would require the 
accurate reproduction of the intricate internal mechanical conditions 
experienced by  the samples used in the experiment. In this particular case, the 
sample strain, where Sn-grown single crystals are taken as a reference, is 
substantial enough to maintain the collapsed-tetragonal phase at the ambient 
pressure, and as a consequence, cannot be ignored in any theoretical description 
pretending to be an accurate portrayal of reality. 

To get around this problem, we chose a different strategy. We exploit the  
linear relationship between the quadrupole splitting and the electric field 
gradient (Eq.~\eqref{eq:quadrupole_splitting_simple}). The proportionality 
constant is the electron charge multiplied by the $^{57}$Fe nuclear quadrupole 
moment, which is a well determined quantity, with values 0.15-0.17 barn 
\cite{Sinnecker2005, Lauer1979, Dufek1995, Martinez2001, Zhang2003, Zhang2003_2, 
Zhi2004}. This allows us to compute the quadrupole splitting by calculating 
$V_{zz}$. We consider unit cell parameters experimentally determined at 
temperature $T$, \cite{ran11a} calculate the quadrupole splitting $\Delta_{\mathrm{DFT}}$ for 
a range of As $z$-coordinates and then perform a polynomial fit of the 
calculated data points to obtain a curve $\Delta_{\mathrm{DFT}} (T,z(T))$, where 
$z(T)$ denotes As $z$-coordinate at temperature $T$. We  then determine the 
$z$-coordinates by solving numerically $\Delta_{\mathrm{DFT}} (T,z(T)) = 
\Delta_{\mathrm{exp}} (T)$, where $\Delta_{\mathrm{exp}} (T)$ is the quadrupole 
splitting experimentally determined at temperature $T$. The As $z$-coordinates 
obtained in this way are shown in Fig. \ref{fig:as_zh} along with the height of 
As atoms above the Fe plane. Although the quadrupole splitting shows 
non-monotonous behavior in the 0-90 K region, reversing the decrease around 30 
K, the As $z$-coordinate, and corresponding height above the Fe plane display a 
monotonous increase and decrease respectively, indicating the well known complex 
relationship between the position of the As atom and the electronic properties 
\cite{Stewart2011}.

\begin{figure}
\begin{center}
\includegraphics{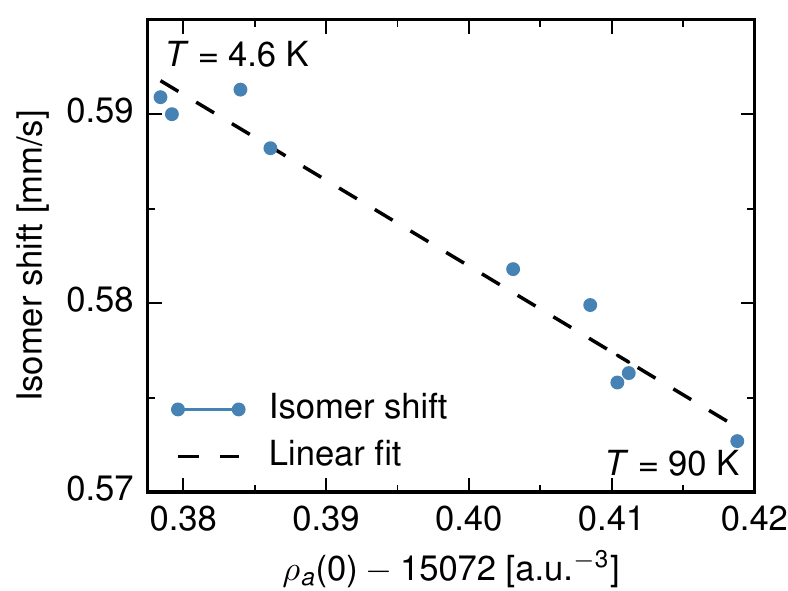}
\end{center}
\caption{(Color online) The linear relationship between the calculated charge 
density at  the iron nucleus and the measured isomer shift in the 
collapsed-tetragonal phase. Note: implicit temperature variation is shown by temperature values for the end points.}\label{fig:is_theo}
\end{figure}

Since the As $z$-coordinates now satisfy  
Eq.~\eqref{eq:quadrupole_splitting_simple} by construction, being able to 
simultaneously satisfy Eq.~\eqref{eq:isomer_shift_simple} provides an important 
consistency check. In Fig. \ref{fig:is_theo} we show the relationship between 
the measured isomer shift and the electron density calculated for the As 
$z$-coordinates obtained from the quadrupole splitting. The data is fitted very 
well by a linear function and the slope is determined to be $a=-0.45\pm0.04$ 
[a.u.$^3$~mm/s] which is in good agreement with previously reported values for 
$^{57}$Fe in various compounds \cite{Lovell2002, Neese2002, Sinnecker2005, 
Han2006}. 

It is important to note here that Figs. \ref{fig:as_zh} and  \ref{fig:is_theo} 
are restricted to the collapsed-tetragonal phase, since in the tetragonal phase 
the experimentally measured values of the quadrupole splittings cannot be 
reproduced by the calculations within a reasonable range of As $z$-coordinate 
values. We attribute this to the fact that antiferromagnetic fluctuations and 
electronic correlation effects  cannot be accurately captured within DFT. In the 
collapsed tetragonal phase it has been shown that both  correlation effects 
\cite{dha14a,Diehl2014, Mandal2014} and  antiferromagnetic fluctuations 
\cite{dha14a,Pratt2009} are suppressed resulting in a good agreement between our DFT 
calculations and the M\"ossbauer measurements.

Focusing now on the tetragonal phase, it is well known that matching the 
experimentaly determinedl and the DFT calculated equilibrium, As $z$-coordinates in Fe-based 
superconductors is, in general, problematic as has been pointed out by numerous 
publications \cite{Yin2008,Mazin2008,Opahle2009,Aichhorn2011}. However, spin 
polarized calculations within GGA have been shown to provide reasonable 
estimates, specially for the orthorhombic phases with long range magnetic order 
\cite{Mazin2008}. 

Such approach can be used here to mimic  the effect of electron correlation and 
antiferromagnetic fluctuations in the tetragonal phase in the framework of DFT. 
We have experimented with spin-polarized GGA calculations with and without 
on-site Coulomb repulsion. In this way we were able to extract more reasonable 
estimates for As $z$-coordinates. Here we present results with $U$=3 eV and 
Hund's coupling $J$=1 eV. We assume N{\'e}el order in order to preserve the 
tetragonal symmetry and account for the antiferromagnetic character of 
fluctuations. In this way we obtain  an  estimate of the As $z$-coordinate in 
the tetragonal phase which {\it should be taken only as an indication of the general 
temperature trend}. An LDA+DMFT (Local Density Approximation plus
Dynamical Mean Field Theory) calculation would be more appropriate but it is 
beyond the scope of the present work. 

\begin{figure}
\begin{center}
\includegraphics{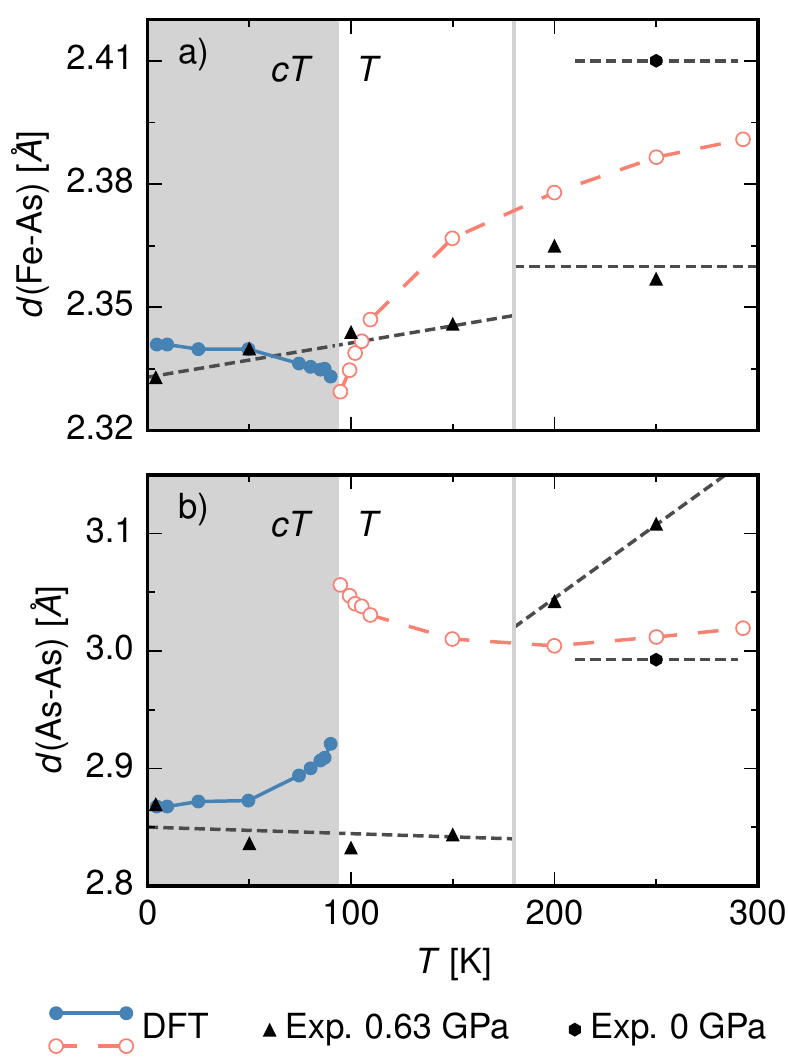}
\end{center}
\caption{(Color online)  (a) The \emph{ab-initio} predicted Fe-As bond length in 
the 960$^{\circ}$ C quenched sample compared against the measured bond lengths 
on Sn-grown samples under hydrostatic pressure \cite{kre08a}. Dotted  lines are 
guides for the eye. (b) Predicted interlayer As-As bond lengths in the 
960$^{\circ}$ C quenched sample compared against the Sn-grown samples 
\cite{kre08a}. Vertical line marks $T_{cT}$ at $P = 0.63$ GPa. Please note that the comparison to experimental data is intented 
to estimate the properties of the ``intrinsic" pressure in 960$^{\circ}$ C 
quenched sample (see the text for details). }\label{fig:bonds}
\end{figure}

Figure \ref{fig:bonds} a) shows the DFT predicted Fe-As bond lengths both in the 
collapsed tetragonal and the tetragonal phase. As the sample is cooled down the 
Fe-As bond length decreases, driven by thermal contraction. Interestingly, this 
trend reverses in the collapsed tetragonal phase where the Fe-As bond length 
increases, working against the thermal contraction of the sample. It is 
instructive to compare this behavior to the experimentally determined Fe-As bond 
lengths at ambient pressure and 0.63 GPa in Sn-grown  samples subjected to 
hydrostatic pressure \cite{kre08a}. We would like to stress here that  the data 
shown corresponds to available measurements~\cite{kre08a} for CaFe$_2$As$_2$ 
performed under \emph{different conditions} to those used in the present study 
and thus our {\it ab initio} DFT results  are not meant to reproduce these data. 
However, as we will show below, comparison to Sn grown, low strain, samples at $P = 0$ and $P = 0.63$ GPa provides references to help define the expected range of behavior seen in our  CaFe$_2$As$_2$  sample.

All measurements in our study are performed at ambient pressure and on samples 
that have a certain amount of strain relative to the Sn-grown samples from Ref.~ 
\onlinecite{kre08a}. We can thus consider the samples used in our study to be 
\emph{intrinsically} under certain amount of pressure, even at zero external 
pressure. Keeping this analogy in mind we can proceed now with the structural 
comparison.  We  observe that our predicted bond lengths fit into the 
experimentally observed range  of bond lengths and that, despite the fact that 
the predicted bond lengths in the tetragonal phase are just a rough estimate, 
they are consistent with the experimental data when we take into account the 
differences in experimental conditions. Namely, the hydrostatic pressure 
stabilizes the cT phase below 200 K at 0.63 GPa in Sn-grown samples, while in 
our $960^{\circ}$ C quenched samples the cT phase is stabilized below 90 K and 
at ambient pressure. Hence, the ambient-pressure $960^{\circ}$ C quenched sample 
structure should roughly correspond to the structure of the Sn-grown sample at 
an intermediate pressure between ambient pressure and 0.63 GPa which is 
precisely what the Fe-As bond lengths in Fig.~\ref{fig:bonds} a) show. The 
exception is a 50 K interval around the T-cT transition which is a consequence 
of hysteresis effects associated with the volume collapse. Since the collapsed 
phase is much stiffer than the tetragonal phase~\cite{Tomic2012, mit11a, 
Yildirim2009} we expect reduction of the structural differences between the 
Sn-grown samples under pressure and the $960^{\circ}$ C quenched samples, which 
is again confirmed by our data. Similar consistency in bond length values as well as jump at $T_{cT}$  is observed in the behavior 
of interlayer As-As bonds, shown in Fig.~\ref{fig:bonds} b).

However, the most striking feature observed in Fig.~\ref{fig:bonds} a) is the 
reversal of Fe-As bond contraction in the cT phase and we believe that this is 
the key to understanding the mechanism behind the stabilization of cT phase at 
ambient pressure. Since, the formation of the cT phase is precipitated by the 
formation of interlayer As-As bonds, we need to look closer into the interplay 
of Fe-As and As-As bonding dynamics (Fig.~\ref{fig:bonds} b)). 

\begin{figure}
 \begin{center}
 \includegraphics{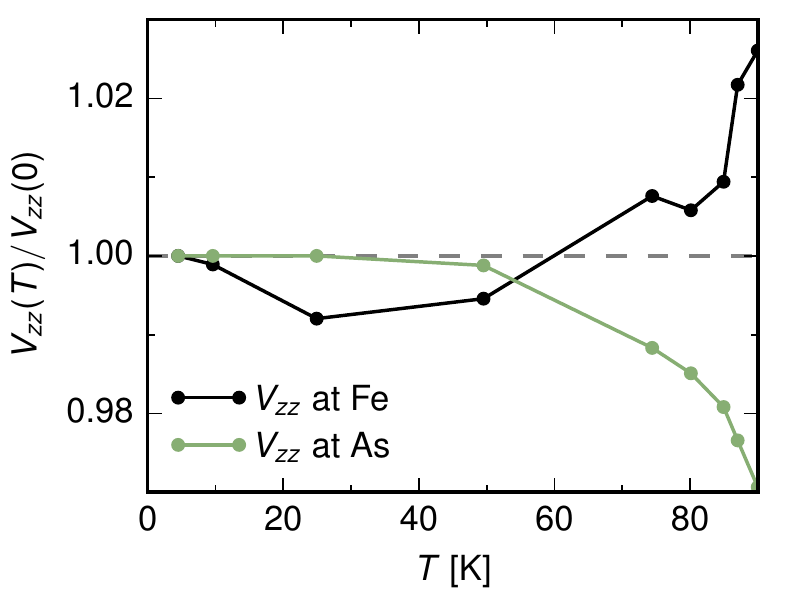}
 \end{center}
\caption{(Color online) The calculated $V_{zz}$ component of the electric field 
gradient for the iron and arsenic sites normalized to their respective zero 
temperature values.}\label{fig:vzz}
\end{figure}

Examination of the behavior of the $V_{zz}$ electric field gradient component 
for the iron and arsenic nuclei (Fig. \ref{fig:vzz}) in the collapsed tetragonal 
phase in CaFe$_2$As$_2$ shows that each component follows a different trend. 
While the electric field gradient decreases at the iron nucleus, it increases at 
the arsenic nucleus as a consequence of the bonding behavior. Relative weakening 
of the Fe-As bond implies that charge density in the bonding region is reduced, 
leading to the reduction of the electric field gradient on the iron site. This 
is reflected in the angular momentum resolved electric field gradient tensor 
\cite{Koch2010} which shows a major contributions coming from the $l=1$ 
component, implicating reduced hybridization of iron $3d$ orbitals with arsenic 
$4p$ orbitals as a main culprit leading to the reduction of the electric field 
gradient. On the other hand, the increase of the electric field gradient at the 
arsenic site follows from the stronger interlayer As-As bonding as the sample is 
cooled. This is also reflected in the angular momentum resolved electric field 
gradient, with $l=1$ component being essentially the only contribution. Such 
bond dynamics indicates that some of the charge in the Fe-As bond region is 
transfered to the interlayer As-As bond region, contributing to its strength and 
helping to stabilize the collapsed tetragonal phase at ambient pressure.

For further answers we examine in Fig.~\ref{fig:bonds} b) the predicted 
interlayer As-As bond lengths for the 960$^{\circ}$ C quenched 
sample at ambient pressure. We again compare with the effect of the hydrostatic 
pressure and show the measured values for Sn-grown CaFe$_2$As$_2$ samples at the 
pressure of 0.63 GPa. We observe that the As-As bonds appear to be relatively 
longer in the tetragonal phase of the 960$^{\circ}$ C quenched sample. Such 
bond configuration can lead to charge saturation in the Fe-As bond region under 
thermal contraction, meaning that at a certain temperature it becomes 
energetically more favorable to transfer some of the charge into the relatively 
less charge dense interlayer As-As bond region, precipitating the formation of 
the interlayer As-As bond which in turn stabilizes the collapsed tetragonal 
phase. This process continues in the collapsed tetragonal phase, where the Fe-As 
bond keeps getting weaker while the interlayer As-As bond gets stronger until 
zero temperature is reached, where the bond lengths revert to the ratio observed 
in the Sn-grown samples under hydrostatic pressure.

\section{Summary}

Experimental clear features in the temperature dependencies of the isomer shift, 
relative spectral area and quadrupole splitting are observed at the transition 
from tetragonal to cT phase in CaFe$_2$As$_2$. From the analysis of the 
temperature dependent isomer shift and spectral area data, as well as the 
temperature evolution of the intensity ratio, an  average stiffening of the 
phonon modes in the cT phase is inferred. Quadrupole splitting increases by 
$\sim 25\%$ on cooling from room temperature to the cT transition, in agreement 
with a large anisotropy of thermal expansion in this temperature region. {A 
clear unusual cusp was observed in the QS(T) data at the cT transition. The 
origin of it is not clear and requires additional studies.}

Additionally, we performed  {\it ab initio} density functional theory 
calculations of the electric field gradient and the electron density  of 
CaFe$_2$As$_2$ in the tetragonal and collapsed tetragonal phase. Comparison of 
these calculations to the measurements allows for the  determination of the As 
$z$-coordinates in the  960$^{\circ}$ C quenched sample, providing the full 
structural information under these conditions and allowing us to probe the 
microscopic origin of the experimental observations. 

We have found that the resulting Fe-As bond lengths are consistent with the data 
recorded for Sn-grown samples under hydrostatic pressure, as expected from the 
broad similarity of the observed structural effects and proper accounting of 
differences in experimental conditions. In the tetragonal phase, as the sample 
is cooled, the Fe-As bonds undergo thermal contraction, resulting in larger 
electric field gradients at the iron nuclei and leading to an increase of the 
quadrupole splitting. In the collapsed tetragonal phase we observe instead an 
anomalous behavior of the Fe-As bond, whereby it undergoes expansion as the 
sample is cooled, resulting in a decrease of the observed quadrupole splitting. 
We attribute this behavior to charge redistribution from the Fe-As bond region 
into the interlayer As-As region, increasing the As-As bonding strength and 
resulting in stabilization of the collapsed tetragonal phase. In our proposed 
scenario, it is high temperature quenching at 960$^\circ$ C which stabilizes the
relatively shorter Fe-As bonds, leading to charge saturation of the Fe-As bond 
region as the sample contracts.

\begin{acknowledgments} 
X.M. was supported in part by the China Scholarship Council. Work at the Ames 
Laboratory was supported by the US Department of Energy, Basic Energy Sciences, 
Division of Materials Sciences and Engineering under Contract No. 
DE-AC02-07CH11358. M.T. and R.V. acknowledge financial support by the DFG 
(Deutsche Forschungsgemeinschaft) through grant SPP 1458.

\end{acknowledgments}

\clearpage

\clearpage

\clearpage

\clearpage

\clearpage

\clearpage

\clearpage

\clearpage

\clearpage

\clearpage

\end{document}